\documentclass[final]{elsarticle}
\usepackage{graphicx}
\usepackage{amssymb,amsmath}
\usepackage{amsthm}
\usepackage{bm}
\usepackage[latin1]{inputenc}

\begin{document}

\title{3D versus 2D domain wall interaction in ideal and rough nanowires}

\begin{frontmatter}

\author{A. Pivano}  \author{Voicu O. Dolocan}\ead{voicu.dolocan@im2np.fr} \address{Aix-Marseille University, Marseille, France}  \address{IM2NP CNRS, Avenue Escadrille Normandie Niemen, 13397 Marseille, France}

\begin{abstract}
The interaction between transverse magnetic domain walls (TDWs) in planar (2D) and cylindrical (3D) nanowires is examined using micromagnetic simulations. We show that in perfect and surface deformed wires the free TDWs behave differently, as the 3D TDWs combine into metastable states with average lifetimes of 300ns depending on roughness, while the 2D TDWs do not due to 2D shape anisotropy. When the 2D and 3D TDWs are pinned at artificial constrictions, they behave similarly as they interact mainly through the dipolar field. This magnetostatic interaction is well described by the point charge model with multipole expansion. In surface deformed wires with artificial constrictions, the interaction becomes more complex as the depinning field decreases and dynamical pinning can lead to local resonances. This can strongly influence the control of TDWs in DW-based devices.

\end{abstract}


\end{frontmatter}


\section{Introduction}

Magnetic domain walls (DW) dynamics in nanowires is an intensive research topic nowadays, with proposed applications ranging from logic to storage devices\cite{Allwood,Parkin,Hayashi}. These devices usually exploit the displacement and the pinning by artificially engineered defects of DWs in planar nanowires (nanostrips).

A large variety of geometrical constrictions and/or imperfection were studied as pinning centers\cite{Klaui,Glathe1,Petit1,Franchin} to manipulate the DWs position. All types of imperfections create a pinning potential for the DW that depends on the DW type (transverse or vortex)\cite{Petit2}. In the case of a transverse wall, the constrictions create a potential well. 

Transverse walls represent the stable state in low width/diameter nanowires. TDW in cylindrical nanowires display different properties than in nanostrips: the Walker limit is completely suppressed, theirs velocity and precession speed depend linearly on the applied current\cite{Yan2} while they propagate without deforming theirs internal structure.

In this paper, we compare the dynamics and the interaction of TDW in ideal and rough cylindrical and planar nanowires. In cylindrical ideal nanowires, the DWs can form metastable states due to the absence of perpendicular anisotropy facilitating the rotation of DWs magnetization in the wall plane\cite{Dolocan3}. Small surface roughness increases the lifetime of the metastable states due to the local pinning. In planar nanowires, the shape anisotropy maintains the DWs magnetization in the plane and no metastable state can form. When the DWs are pinned at symmetric notches, they interact mainly through the dipolar field in both types of wires, and this interaction can be well described by the point charge model\cite{Kruger}. In surface deformed wires, the depinning field decreases and DW resonances can occur due to the pinning landscape. The influence of the temperature is discussed employing the stochastic 1D model. Our results show the importance of understanding the pinning mechanism and of mutual interaction between DWs in nanowires.


\begin{figure}[b!]
  \includegraphics[width=8cm]{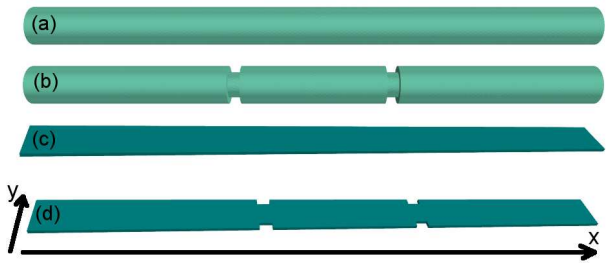}\\
 \caption{\label{Fig.1} (Color online) Simulated structures: (a) perfect cylinder, (b) cylinder with two radial constrictions, (c) perfect planar wire and (d) planar wire with two symmetric double notches.}
\end{figure}

\section{Model and geometry}

We computed a large number of 3D micromagnetic simulations using the nmag package\cite{Fischbacher} to determine the dynamics of DWs in nanowires\cite{Dolocan1}. Fig.~\ref{Fig.1} shows the different simulated structures: cylindrical wires with length of 900nm or 1200nm and 60nm in diameter and planar wires with same lengths (900nm, 1200nm) and a width of 60nm and 5nm thickness. The wires were discretized into a mesh with a cell size of 4nm, inferior to the exchange length ($\sim$5nm for Ni). We use the material parameters of Nickel: saturation magnetization $\mu_0$M$_{s}$=0.6T, exchange stiffness A = 1.05 $\times$ 10$^{-11}$J/m, $\gamma$=188.5GHz/T (g factor of 2.15), damping parameter $\alpha$=0.015 \cite{Gilmore,Heinrich}. No magnetocrystalline anisotropy is considered and the temperature is set to T = 0K. The stable structure in our nanowires is a V-shaped TDW separating domains pointing along the long axis ($x$ axis). The average position of each DW was extracted along with the azimuthal angle $\psi$ of the magnetization inside the DW. 

The constrictions in both types of wires have the same length (20nm) and depth (10nm). In the cylindrical wire they were created as a sharp modulation of the nanowires diameter, while in the planar wire the usual symmetric rectangular notches are used. The surface deformation was created from a random profile along the $x$ axis in both cases, with a predefined amplitude deformation $\sigma$ and variable correlation length $\lambda$. The profile is symmetric on the upper and lower edges of the planar wire\cite{Nakatani,Martinez1} and full revolved around $x$-axis for the cylindrical wire.


\section{Numerical results}

\subsection{Perfect wires}


\begin{figure}[t!]
  \includegraphics[width=8cm]{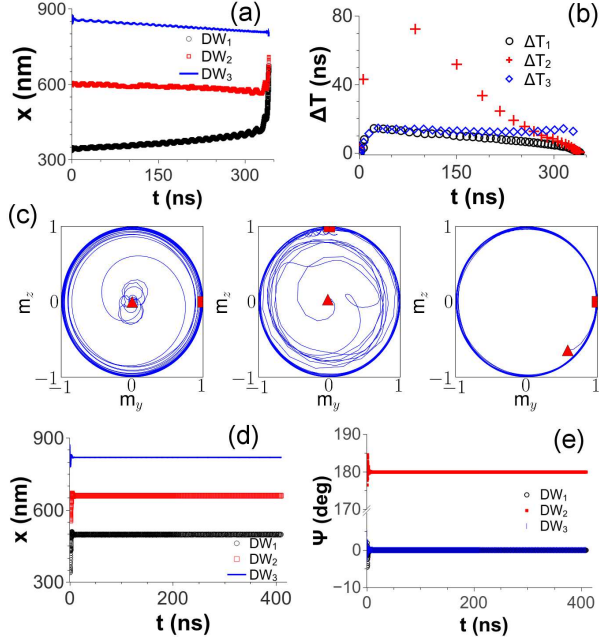}\\
 \caption{\label{Fig.2} (Color online)  Interaction between three DWs in a perfect Nickel nanowire (length = 1200nm, diameter = 60nm): (a) Spatial displacement of DWs during 350ns in a cylindrical wire. (b) Synchronization of two DWs from panel (a) determined by the instantaneous period. (c) Rotation of the magnetization in the $yz$ plane at the DW center for each DW from panel (a). The squares denote the initial position, while the triangles represent the final position. (d) Spatial displacement of DWs and (e) precession angle $\psi$ during 400ns in a planar wire.}
\end{figure}

We start with the study of the interaction between DWs in the same ideal nanowire in the absence of pinning. The comparison between the two types of wires is detailed in Fig.~\ref{Fig.2}. The simulation is initialized with three DWs in the nanowire (length 1200nm, diameter/width 60nm), with exterior DWs pointing in the same direction and central one pointing at 90$^{\circ}$ (cylinder) or 180$^{\circ}$ (strip), that are free to relax. In the cylindrical wire (panels (a)-(c)), the magnetization in the DW can precess freely around the longitudinal $x$-axis as no perpendicular anisotropy is present. The dynamics in the wire is dominated by the interaction between the DWs. The exterior DWs also interact with the ends of the nanowire\cite{Dolocan3}, but this interaction is small as the distance between DW and the ends is at least of 340nm, superior to the initial distance between DWs (260nm). As initially the DWs are distanced by more than 200nm, the main contribution to the interaction comes from dipolar energy which tries to keep their magnetization direction antiparallel. The middle DW magnetization rotates to keep it antiparallel with the exterior DWs. As they precess, the DWs also oscillate spatially along $x$-axis forming a metastable state (described in detail in \cite{Dolocan3}). After 280ns, the left and middle DWs synchronize as can be determined from their instantaneous period calculated from the Hilbert transform. The synchronization is a result of the exchange energy which becomes the dominant factor in the interaction energy at close range. When the DWs become parallel, they can annihilate (320ns) as the exchange energy drops abruptly and the dipolar energy remains almost constant. As long as the magnetization in the two DWs is antiparallel, they cannot annihilate due to the important exchange energy at close separation. This can be observed in panel (d) and (e) for planar nanowires with consecutive DWs antiparallel. The DWs cannot annihilate freely as the magnetization is frustrated (by the anisotropy) and cannot turn unimpeded in the $yz$ plane. Initially, the DWs attract due to the dipolar field and they approach each other, but rapidly after 2ns they repel and remain stationary at a fixed distance. The $\psi$ angle (panel (e)) stays along the $y$ direction having a small initial oscillation ($\sim 3^{\circ}$). If the DWs are initiated with the same direction of magnetization in planar nanowires, they annihilate rapidly in less than 1ns (image not shown)\cite{Kunz}.

\subsection{Rough wires}


\begin{figure}[t!]
  \includegraphics[width=8cm]{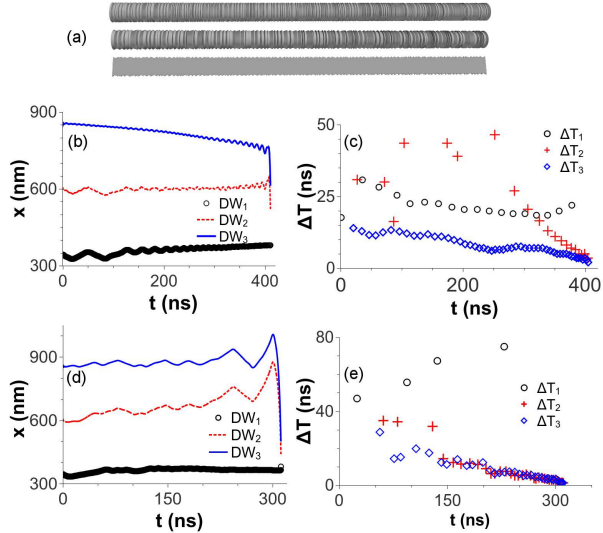}\\
 \caption{\label{Fig.3} (Color online) Meshes of the surface deformed wires used in simulations (a). Spatial displacement of DWs and theirs instantaneous period for a cylindrical nanowire with radial deformed surface as in (a): (b) and (c) correspond to soft roughness with amplitude deformation $\sigma$=2nm (upper wire in (a)) while (d) and (e) to hard roughness ($\sigma$=4nm).}
\end{figure}

Usually, nanowires always present some kind of surface roughness. The cylindrical wires were radially deformed while the planar wires were edge deformed. Two deformation amplitudes were used: $\sigma$=2nm with $\lambda$=3-10nm and $\sigma$=4nm with $\lambda$=5-15nm. An example of meshes used in simulations are shown in Fig.~\ref{Fig.3}(a).

To compare with the ideal case we used the same initial configuration of three DWs situated at equal distance. Five different meshes were studied for each deformation amplitude. We observe two different behaviors in cylindrical wires: at small deformation (as in Fig.~\ref{Fig.3}(b) and (c)), the DWs sometimes effectuate small jumps depending on the local pinning profile and annihilate after a longer time than in the smooth case (around 400ns) for four cases out of five. In the case presented in panels (b) and (c), the left and middle DW rotate clockwise initially. After 100ns, the middle DW starts to change its rotation direction and to rotate in the same direction as the right DW (anticlockwise rotation). It oscillates around a fixed position with a small amplitude (8nm) for almost 200ns, while the other two DWs slowly move closer. The right DW moves closer to the middle DW (which stays pinned) compared to the left DW and therefore it interacts stronger (the angular velocity of the DW is proportional with the magnetostatic field\cite{Yan2}). The magnetization in the middle DW starts to oscillate between clockwise and anticlockwise direction until finally stabilizes and starts to synchronize with the right DW. When it finally synchronizes, after 380ns, it moves away form its local trapping position and annihilates. At higher deformation (Fig.~\ref{Fig.3}(d) and (e)), the DWs sometimes effectuate larger jumps and annihilate after an identical or lower time than the ideal case (around 300ns) for four cases of five trials. In the presented case, the left and middle DWs rotate clockwise, but rapidly after 50ns, the middle DW starts to change its rotation direction as it jumps closer to the right DW while the lower DW remains pinned at a local position. After 150ns, it oscillates almost synchronously with the right DW but it does not annihilate until 320ns due to the larger spatial jumps. Local pinning sites trap the DWs and a larger interaction energy is needed to depin them hindering the annihilation. This local pinning also influences the interaction between DW pairs, some DWs can approach while others stay pinned and therefore can change theirs rotation direction. 

The edge roughness does not influence the interaction of DWs in planar nanowires, as the DWs magnetization is still trapped in-plane. Therefore, no metastable states appear, with or without surface deformation.

\subsection{Pinning at artificial constrictions in a perfect wire}



\begin{figure}[t!]
  \includegraphics[width=6cm]{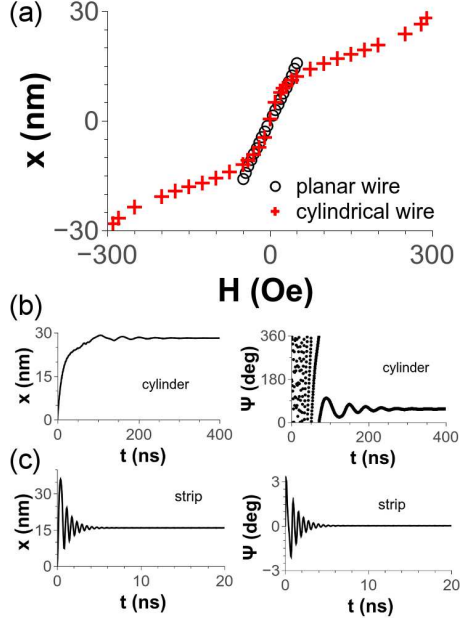}\\
 \caption{\label{Fig.4} (Color online) (a) Equilibrium position for a pinned DW function of the applied dc magnetic field for a planar or cylindrical wire with a symmetric constriction at its center. The oscillations of the spatial position and azimuthal angle of the DW are shown for a dc field of 290Oe for a cylindrical wire in (b) and for a dc field of 49Oe for a nanostrip in (c).}
\end{figure}

To be able to compare the DWs interaction between the two types of wires, we examined the case when the DWs are pinned at artificial constrictions in a smooth nanowire. For the planar wires we used two symmetric notches as a pinning center, while for the cylindrical wires we used a radial constriction (as shown in Fig.~\ref{Fig.1}).

We first evaluate the static interaction of a DW with a single pinning center, situated in the center of the wire, from quasistatic micromagnetic simulations. The average equilibrium position is shown in Fig.~\ref{Fig.4}(a) for both wire types. For planar wires two symmetric rectangular notches create a harmonic potential ($x$ linear in $H$), while a radial notch in a cylinder creates a complex potential with a change of slope when the DW center is pinned outside the notch\cite{Dolocan4}. The calculated spring constants and depinning fields are $k=8\times 10^{-5}$N/m and H$_{dep}$=50Oe for the planar wire, $k_1=1.23\times 10^{-3}$N/m and $k_2=3.21\times 10^{-4}$N/m and H$_{dep}$=300Oe for the cylindrical one. The radial notch has a larger depinning field and the DW relaxes to equilibrium doing several rotations in the $yz$ plane (Fig.~\ref{Fig.4}(b)), however the DW pinned at a radial notch has also a very long relaxation time (hundreds of ns) compared to the symmetric rectangular notches (Fig.~\ref{Fig.4}(c)).

The domain wall width suffers a small modification when submitted to the external field. The width contracts by 2 nm just before depinning from 20nm (without field) to 18nm for the cylindrical wire and from 31.9nm to 30.9nm for the nanostrip.

\begin{figure*}[bt!]
  \includegraphics[width=14cm]{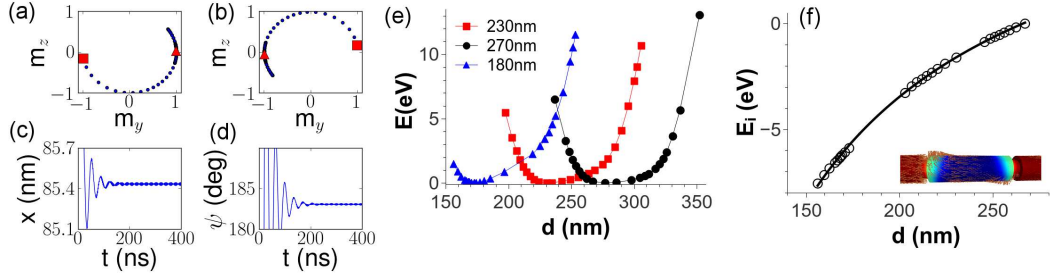}\\
 \caption{\label{Fig.5} (Color online) Interaction between two DWs pinned at notches: (a) and (b) Rotation of the magnetization in the $yz$ plane at the DW center for two DWs pinned at notches separated by 180nm, when an external field of 10Oe was applied. The squares denote the initial position, while the triangles represent the final position. (c) and (d) Relaxation of the DW position and precession angle for  the DW from (b). (e) Variation of the normalized potential energy of the pinned DWs with the separation distance between them for three notch separation of 180, 230 and 280nm. (f) Variation of the interaction energy between the two DW with the separation distance (circles) for the same three notch separation as in (e). The line represents a fit with the multipole expansion. The inset shows two DWs pinned at notches separated by 180nm.}
\end{figure*}

We analyze next the DWs interaction considering two DWs of opposite chirality head-to-head and tail-to-tail (HHDW and TTDW) pinned at notches situated in the same wire. We start with the cylindrical wire, where several edge-to-edge distances between notches were used: 180nm, 230nm and 270nm, with identical notch form. The DWs interact through the dipolar field and theirs equilibrium positions in absence of external perturbation is not at the center of each notch. They precess around the cylinder axis and relax with the magnetization pointing in opposite directions at theirs center, corresponding to an attractive interaction, to minimize energy. An example of the interaction between the two DWs is shown in Fig.~\ref{Fig.5}(a)-(b) for the 180nm separation distance, where the orientation of the DWs center magnetization in the $yz$ plane is shown. An external magnetic field of +10Oe is applied to push closer the DWs. Theirs initial orientation, indicated by squares, is opposite and after 200ns they relax to equilibrium maintaining an opposite orientation of magnetization (final position indicated by triangles). In panels (c) and (d), the relaxation process is shown for the right DW and a very small oscillation around the equilibrium position is visible after the 200ns due to the local pinning potential as the center of the DW is situated at the edge of the notch. The closest position of the two DW before depinning is illustrated in the inset of the panel (f) for an applied field of +125Oe. 

Due to their interaction, the equilibrium position of each DW is closer to the other notch when the notches are close together, as can be observed in Fig.~\ref{Fig.5}(e) where is represented the potential profile variation with the distance between the two DWs. The minimum of the potential well (no external perturbation present) is outside of the notches for the closest separation (180nm), when the DWs are situated at a distance of 13.5nm from the center of each notch. The pinning potential is highly distorted in this case as more field is needed to separate them. The interaction between the DWs also greatly changes the depinning field for the two directions of the nanowire. 


To quantify the interaction between the two DWs pinned at notches, we represent in Fig.~\ref{Fig.5}(f) the variation of the interaction energy between the DWs with their separating distance. The interaction energy was calculated as follows: first the energy of the nanowire without any DW was calculated (edge energy). Then, the pinning energy for each DW was calculated similarly with the case of a single notch: we initialized the nanowire with a single pinned DW (left or right) and determined the potential profile of each of the notches. Afterwards, we extracted from the total energy the pinning energy of each DW and the initial demagnetizing energy (edge energy), to obtain the interaction energy. In the figure, we represented only the interaction energy (normalized) when pushing the DWs closer for each of the three notch separations studied. As observed from the fitted curve, the calculated points fall on an universal curve which correspond to the dipolar interaction (multipole expansion) as discussed below. 

The interaction energy between two DWs of opposite chirality in the same wire was calculated using the point-charge model with multipole expansion\cite{Kruger}. The expression disregards the internal structure of the DWs and the exchange interaction between them (thus applicable at high separation of DWs), considering only the interaction through the dipolar field therefore giving a qualitative description of the interaction. The interaction energy between two DWs of opposite chirality (HHDW and TTDW) separated by $d$ is:

\begin{align}
\label{eq1}
E &= \frac{\mu_0q_1q_2}{4\pi d} \left( 1 - \frac{\pi^2\Delta^2\cos(\psi_1-\psi_2))}{4d^2} \right) +E_{Qq} \nonumber\\ &= -\frac{\mu_0M_s^2S^2}{\pi d}\left( 1 - \frac{\pi^2\Delta^2\cos(\psi_1-\psi_2))}{4d^2} + \frac{2\pi^2\Delta^2 - w^2 - t^2}{12d^2} \right)
\end{align}

\noindent where $q_{ij}$ represent the 'magnetic charges' of the DWs and in this approximation are equal to $\pm2M_s S$, with $S$ the section of the nanowire and $\Delta$ the DW width. The first term represents the Coulomb-type interaction energy (monopole-monopole), while the second is the correction from the dipole-dipole interaction. The third term represents the quadrupole-monopole contribution with $w$ the width and $t$ the thickness of the nanowire. We fitted the above expression to the numerical values of Fig.~\ref{Fig.5}(f) (line). From the parameters of the fitted curve, we extracted the domain wall width and obtained a value of 23.28nm, larger than the value obtained by fitting the profiles of DWs with the Bloch profile which varies between 17.91nm and 19.67nm depending on the applied field and local pinning position.  This discrepancy between the two models comes mainly from the error in estimating the average position of the DW in the 3D wire as the DW distorts taking a bow shape when pinned at the notch while submitted to an external field. This general dependence of the interaction of DWs describes the importance of the dipolar energy when the DWs are situated at large distances, where the exchange energy is much smaller (here by a factor of five).


\begin{figure}[t!]
  \includegraphics[width=6cm]{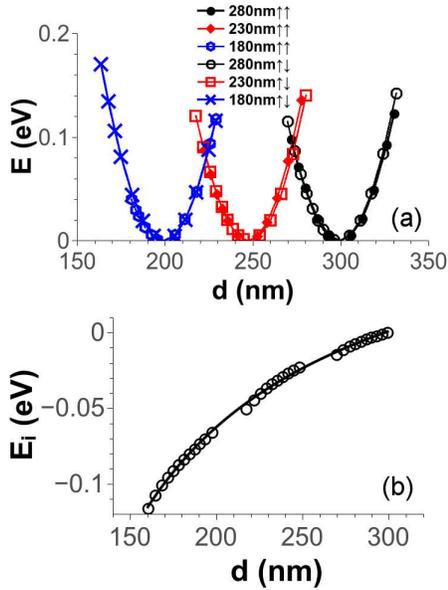}\\
 \caption{\label{Fig.6} (Color online)  (a) Variation of the normalized potential energy of the pinned DWs in nanostrips with the separation distance between them for three notch separation of 180, 230 and 280nm. Two cases for each separation are shown: DWs magnetization parallel or antiparallel. (b) Plot of the interaction energy between two pinned antiparallel DWs.}
\end{figure}

The same analysis as above was effectuated for two DWs of opposite chirality pinned at symmetric rectangular notches inside the same nanostrip (Fig.~\ref{Fig.1}(d)). The DW interaction was studied for several distances between notches (180nm, 230nm and 280nm). As the magnetization of the two DWs is confined to the $y$ direction, two cases were studied: parallel magnetization of the DWs or antiparallel. The potential energy of the pinned DWs varies with the separation distance and is shown in Fig.~\ref{Fig.6}(a) for three notch separations.

When the DWs are pinned at larger distances (230nm, 280nm), the potential energy is quasi identical for parallel or antiparallel DWs. The most important difference appear at 180nm notch separation. When the DWs are parallel, a smaller positive depinning field (DWs pushed closer) is needed as they attract each other and after depinning will annihilate. When the DWs are antiparallel, the positive depinnig field is larger as the DWs repel each other due to the large exchange energy at close separation. After depinning the DWs don't annihilate but form a 360$^{\circ}$ DW\cite{Kunz}. The anisotropy impedes the magnetization to rotate inside the DWs and the antiparallel DWs cannot annihilate. The interaction energy between DWs was extracted in the same manner as for cylinders and is shown in Fig.~\ref{Fig.6}(b). A DW width of 29.30nm was determined from the point charge model, that is closer to the calculated Bloch width (30.96nm - 31.95nm) than in the 3D case. In 2D wires, the error when extracting the DW average position is lower allowing for a better agreement between the calculated DW width by the different models.

\subsection{Pinning at artificial constrictions in a rough wire}


\begin{figure*}[t!]
  \includegraphics[width=14cm]{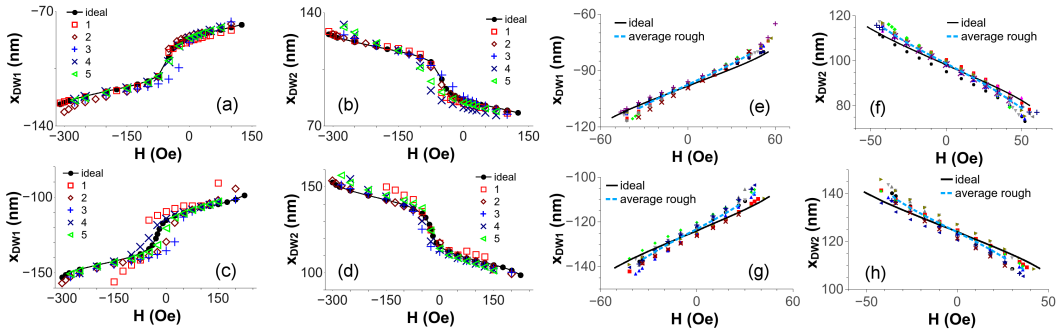}\\
 \caption{\label{Fig.7} (Color online) Variation of the equilibrium position, with applied field, of two pinned DWs for 180nm notch separation in a cylindrical wire (a)-(b) and respectively in a planar wire (e)-(f). The ideal case is compared with several surface deformed cases as explained in the text. The variation of the equilibrium position for 230nm notch separation is shown in (c)-(d) for the cylindrical wire and in (g)-(h) for the planar wire.}
\end{figure*}

The roughness modifies the interaction between DWs pinned at notches. Several simulations were computed for surface deformed wires with $\sigma$=4nm and $\lambda$=3-15nm, using radial deformation for cylinders (5 cases) or edge deformation (10 cases) for strips. The results are summarized in Fig.~\ref{Fig.7}. The variation of the equilibrium position of pinned DWs with applied field for 180nm and 230nm notch separation is depicted for both wires in the ideal and rough cases.

For the deformed cylindrical wire with DWs pinned at 180nm notch separation ((a) and (b)), at positive applied field (DWs pushed closer), the dipolar interaction is strongest and prevails the local pinning due to roughness. The position of the DWs follows almost the same curve as in the ideal case outside of the notch (same slope). For negative applied field, the DW travels through the notch and small variations to the ideal case can appear inside the notch and at large fields. At 230nm notch separation ((c) and (d)), the influence of the surface roughness is more pronounced and dynamical pinning is more easily observable. The depinning field decreases in all deformed wire simulations compared to the ideal wire, when the DWs are pushed closer (positive field), by 25-35Oe at 180nm notch separation and by 25-50Oe at 230notch separation. It also decreases in four out of five cases for negative applied fields with same amounts as for positive fields. In one case, for each notch separation, the depinning field was identical as the ideal one. In terms of the variation of normalized potential energy with the separation distance between DWs (as in Fig.~\ref{Fig.5}(e)), the DWs depin more rapidly when pushed together and the height of the pinning potential well by the notch is diminished. The potential well remains asymmetric with a larger deviation from the ideal case when the DWs are pinned inside each notch. The interaction energy between DWs remains magnetostatic, but a quantitative determination is more complicated due to the lack of precise determination of the pinning energy due to roughness. However, the opposite analysis can be conducted and the pinning energy due to roughness can be determined quantitatively considering that the interaction energy and the pinning energy by a notch are completely determined from the ideal case.  

In deformed nanostrips, a change of slope is observed in eight out of ten simulations, compared to the ideal nanostrip. The average position of the DWs of all ten rough nanostrips is shown, along with all individual cases, in panels (e) to (h) for both notch separation. The average curve (dotted line) has the same slope as eight of the rough nanostrips. The two remaining simulations follow the ideal case or have a different slope. At 180nm notch separation the slope variation is lower than for the 230nm separation, as in cylinders. The depinning field decreases in all cases for 230nm case and for negative fields for 180nm case with 8-14Oe. For the positive applied fields at 180nm separation, the depinning field decreases in six out of ten cases and in the remaining four increases slightly with maximum 6Oe.

\begin{figure}[!]
  \includegraphics[width=7.5cm]{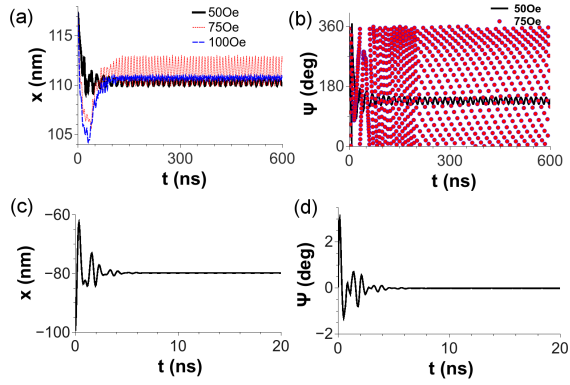}\\
 \caption{\label{Fig.8} (Color online) (a) Plot of the right DW position and (b) azimuthal angle in a deformed cylindrical wire, for different applied fields, showing local resonance. (c) Plot of the DW position and (d) azimuthal angle in a deformed planar wire for $H$ = 42Oe.}
\end{figure}

In the case presented in Fig.~\ref{Fig.8}(a) for a deformed cylindrical wire, the right DW is pinned at the same equilibrium position situated 15nm from the center of the right notch (at 125nm) when the applied field is varied between 50Oe and 125Oe. As can be observed in panels (a) and (b), around 75Oe there is a resonance of the right DW in the local potential, as the DW position oscillates with high amplitude and the magnetization precesses in a steady state around the $x$ axis. At lower fields, the magnetization only precesses initially and afterwards the oscillation damps to the equilibrium position. At resonance, a steady state develops similar to an applied polarized current when the spin torque counter balances the damping. The DW resonance depends on the local potential curvature and therefore on the geometry of the deformation (the deformation profile is close to triangular). This dc magnetic induced resonance resembles the current induced resonance of geometrically confined DWs in rings\cite{Bedau} and in triangular notches in stripes\cite{Bisig}, although without the transformation in the DW structure. The DW stays pinned  until 125Oe, although the spatial oscillation of the DW position diminishes. If the field is increased above the local depinning field (150Oe), the DW makes a jump of 10 to 15nm and behave like in the perfect wire relaxing to a local equilibrium. The resonance phenomena in rough cylindrical wires is observed in all the computations for one or both pinned DWs.

No resonance is observed in the deformed nanostrips, only an asymmetry in the oscillation when relaxing to the equilibrium position as in panels (c) and (d) of Fig.~\ref{Fig.8}.


\section{Discussion and Conclusion}

The interaction energy between DWs in the same nanowire was shown to be mainly dipolar and depend on the roughness. However, the DWs interaction depends also on temperature as the pinning of a DW by a constriction in a nanowire can be greatly influenced by it. This effect is often estimated using the stochastic one dimensional model which assumes the DW as rigid and is usually determined from\cite{Lucassen,Martinez}:

\begin{align}
\label{eq2}
(1+\alpha^2)\dot{X} =& -\frac{\alpha\gamma\Delta}{2\mu_0M_s S}\frac{\partial E}{\partial X} + \frac{\gamma\Delta}{2}H_k\sin 2\psi \nonumber\\
& + \frac{\gamma}{2\mu_0M_s S}\frac{\partial E}{\partial\psi} + \eta_{\psi} - \alpha\eta_X\\ 
(1+\alpha^2)\dot{\psi} =& -\frac{\gamma}{2\mu_0M_s S}\frac{\partial E}{\partial X} -\frac{\gamma\alpha}{2}H_k\sin 2\psi \nonumber\\
& - \frac{\alpha\gamma}{2\Delta\mu_0 M_s S}\frac{\partial E}{\partial\psi} + \eta_X + \alpha\eta_{\psi}
\end{align}

\noindent with $X$ and $\psi$ the position and respectively the azimuthal angle of the DW, $H_k$ the DW demagnetizing field, $\eta_X$ and $\eta_{\psi}$ represent stochastic Gaussian noise with zero mean value and correlations $\langle \eta_i(t) \eta_j(t') \rangle = (2\alpha k_B T)/(\mu_0M_s\Delta S)\delta_{ij}\delta(t-t')$. $E$ is the potential energy of the DW that includes the internal energy, the Zeeman energy, the interaction energy with other DW and the pinning energy.

We evaluated the depinning probability of a DW from a pinning potential corresponding to the single radial notch in the cylindrical nanowire or to the symmetric rectangular notches in nanostrips. The same parameters as in the micromagnetic simulation were used and 1000 realizations were computed for each applied field. In the nanocylinders, at T=0K, the depinning occurs at 299Oe, while at T=300K the depinning probability changes from 0\% to 100\% in a field range of 14Oe (283Oe to 297Oe). In planar wires, at T=0K, the depinning takes place at 50Oe,  while at T=300K the depinning is of 3\% at an applied field of 25Oe. Based on this analysis, the thermal fluctuations shoud impact more the pinning of a DW in nanostrips. This can be deduced also from the height of the pinning potential, which is less than 0.5eV for notches in nanostrips compared with several eV for the radial notch in nanocylinders.

The cylindrical nanowires have advantages but also drawbacks for practical applications as compared to nanostrips. Cylindrical constrictions give much longer DW relaxation time, which increases with roughness. However, as no Walker limit is present, the study of DW interactions in such structures could be the basis for in-depth exploring of fundamental interactions. We speculate that triangular constrictions will lead to interesting resonance states in an applied dc magnetic field as seen in the rough wires.  

In conclusion, we have presented a systematic study of the interaction between TDWs in planar and cylindrical nanowires. Free DWs form metastable states in free anisotropy wires (as in the cylindrical geometry). Two pinned DWs interact strongly through the dipolar field in both wire types. Surface roughness modifies the lifetime of metastable states and the interaction of pinned DWs becomes more complex due to the asymmetry, dynamical pinning and temperature.


The authors wish to thank L. Raymond and is grateful for the support of the NANOMAG platform by FEDER and Ville de Marseille. The computations were performed at the Mesocentre d'Aix-Marseille University.


\end{document}